# From the dynamic to the static glass transition via hypersonic measurements using Brillouin spectroscopy

Jan-Kristian Krüger[1*], Rafael J. Jiménez Riobóo[2], Bernd Wetzel[1], Andreas Klingler[1+]

[1]Leibniz-Institut für Verbundwerkstoffe (IVW), 67655 Kaiserslautern, Germany
[2]Instituto de Ciencia de Materiales de Madrid (CSIC), 28049 Madrid, Spain

**ABSTRACT**. For the fragile, low-molecular-weight liquid diglycidyl ether of bisphenol A (DGEBA), we report on the dynamic glass transition and a further acoustic anomaly in the vicinity of the thermal glass transition based on hypersonic investigations of the longitudinal elastic modulus using Brillouin spectroscopy. This additional acoustic anomaly of the longitudinally polarized phonon is confirmed by the occurrence of an anomaly of the shear phonon at the same thermal glass transition temperature. Analysis of the generalized Cauchy relation suggests that both anomalies are coupled to a glass transition phenomenon independent of the so-called $\alpha$-relaxation process.

## I. INTRODUCTION

The thermal glass transition (TGT) plays a central role in condensed matter physics and technology. This is particularly true for the material class of polymers. Despite its central importance, the nature of TGT is still not fully understood (e.g., [1–13]).

Among the unresolved central questions are: i. Is there a phase transition from the fluid high-temperature phase to the glassy low-temperature phase, and ii. how does the dynamics of glass formation, the so-called alpha relaxation process, control TGT?

A central problem in interpreting experimental data intended to clarify the nature of TGT is often the lack of clear boundary conditions. In particular, when investigating TGTs, temperature changes are often generated by temperature rates, which produce uncontrolled additional dynamic property changes. In this case, measured glass transition temperatures $T_g$, including their kinetics (e.g., [11,14,15]) as well as phase transition properties (e.g., [16–19]), naturally overlap strongly in the vicinity of $T_g$.

In order to eliminate or at least significantly reduce the problem of additional dynamic influences on the TGT caused by temperature rates, we have introduced the temperature jump method (Fig. 1) in the field of Brillouin spectroscopy (e.g., [5,20]) as well as in other measurement methods [21,22]. As $T_g$ is approached, it can take days after a temperature jump until the equilibrium property - in this case the relaxed phonon frequency $f_r$ - is reached.

Based on this approach, we here provide evidence for the existence of an independent static glass transition phenomenon that complements the dynamic glass transition properties.

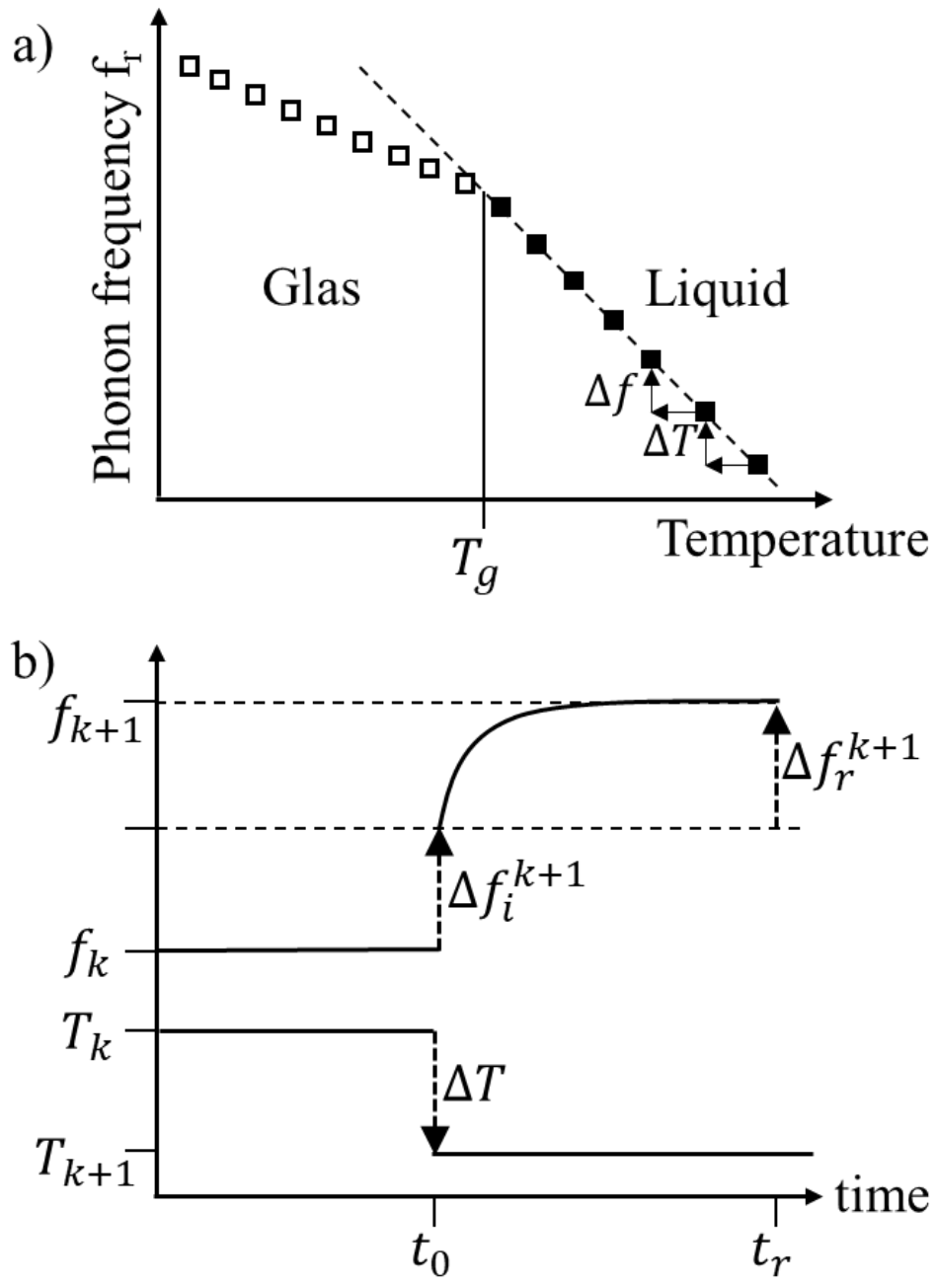

FIG 1. Schematic representations of (a) the equilibrated phonon frequency $f_r$ as function of temperature $T$, (b) the $k+1$ temperature jump ΔT on cooling and the resulting temporal response in the hypersonic frequency range $f$. $t$ is the time, $t_0$ indicates the time of the temperature jump, the indices $i$ and $r$ stand for the instantaneous and the relaxing part of the frequency response, respectively.

## II. METHODS

### A. Material

In this work, we chose a non-crystallizable diglycidyl ether of bisphenol A (DGEBA, D.E.R. 331) as a low-molecular-weight ($M_w = 370\ g/mol$) glass-forming

*Contact author: jan-kristian.krueger@ivw.uni-kl.de
+Contact author: andreas.klingler@ivw.uni-kl.de

model system. The thermal glass transition temperature is approximately $T_g \sim 247\ K$.

### B. Brillouin spectroscopy

For Brillouin spectroscopic investigations, the liquid DGEBA was poured into a special optical cuvette. To prevent adhesion problems between the DGEBA sample and the cuvette windows at lower temperatures, the glass windows of the sample cuvette were coated with a very thin layer of Teflon using friction.

A 6-pass Sandercock-type spectrometer was used as the Brillouin spectrometer [20]. In order to avoid any additional influence of the optical refractive index of the sample on the acoustic wavelength, the 90A scattering geometry introduced by one of the authors [23] was used. If the wavelength of the exciting laser is $\lambda_0 (= 532\ nm)$, the related acoustic wavelength is obtained in the case of the 90A scattering geometry as a constant:

$$\Lambda^{90A} = \lambda_0/\sqrt{2} \quad (1)$$

The acoustic wavelength $\Lambda^{90A}$ and the measured Brillouin frequencies of the associated longitudinal and transverse phonon modes $f_L$ and $f_T$ are used to obtain the associated sound velocities of the isotropic state

$$v_{L,T} = \Lambda^{90A} \cdot f_{L,T} \quad (2)$$

The sound velocities $v_{L,T}$ and the mass density $\rho$ can now be used to determine the associated longitudinal and transverse elastic moduli of the isotropic state

$$c_{L,T} = \rho \cdot v_{L,T}^2 \quad (3)$$

The above equations apply within the framework of linear response theory in all cases under thermodynamic equilibrium conditions. If irreversible thermodynamics play a role [24], the equations only apply to the two limiting cases: internal equilibrium and clamped equilibrium. In the latter case, Eq. 3 reads

$$c_{L,T}^{\infty} = \rho \cdot \left(v_{L,T}^{\infty}\right)^2 \quad (4a)$$

$$v_{L,T}^{\infty} = \sqrt{\frac{c_{L,T}^{\infty}}{\rho}} \quad (4b)$$

## III. RESULTS & DISCUSSION

Fig. 2 shows sound velocities as measured by Brillouin spectroscopy for the longitudinal and transverse polarized acoustic phonon modes in the range of the thermal glass transition (TGT), i.e. in the range of the quasi-static glass transition at $T_g$. Due to the use of the temperature jump method, this measurement took more than 10 days. Since the corresponding sound velocities were measured at frequencies in the GHz range, it can be assumed that the phonon modes are frequency-clamped and therefore that in the vicinity of $T_g$ the clamped sound velocities $v_{L,T}^{\infty}$ were measured.

This raises the interesting question of what phenomenon causes the kink in the sound velocities at $T_g$ if the dynamics responsible for the glass transition process ($\alpha$-process) are already clamped at much higher temperatures. In other words, the kink in the sound velocities $v_L$ and $v_T$, or the associated sound frequencies $f_L$ and $f_T$, cannot originate from the measurement frequency nor from a temperature rate, as the latter was eliminated by the temperature jump method.

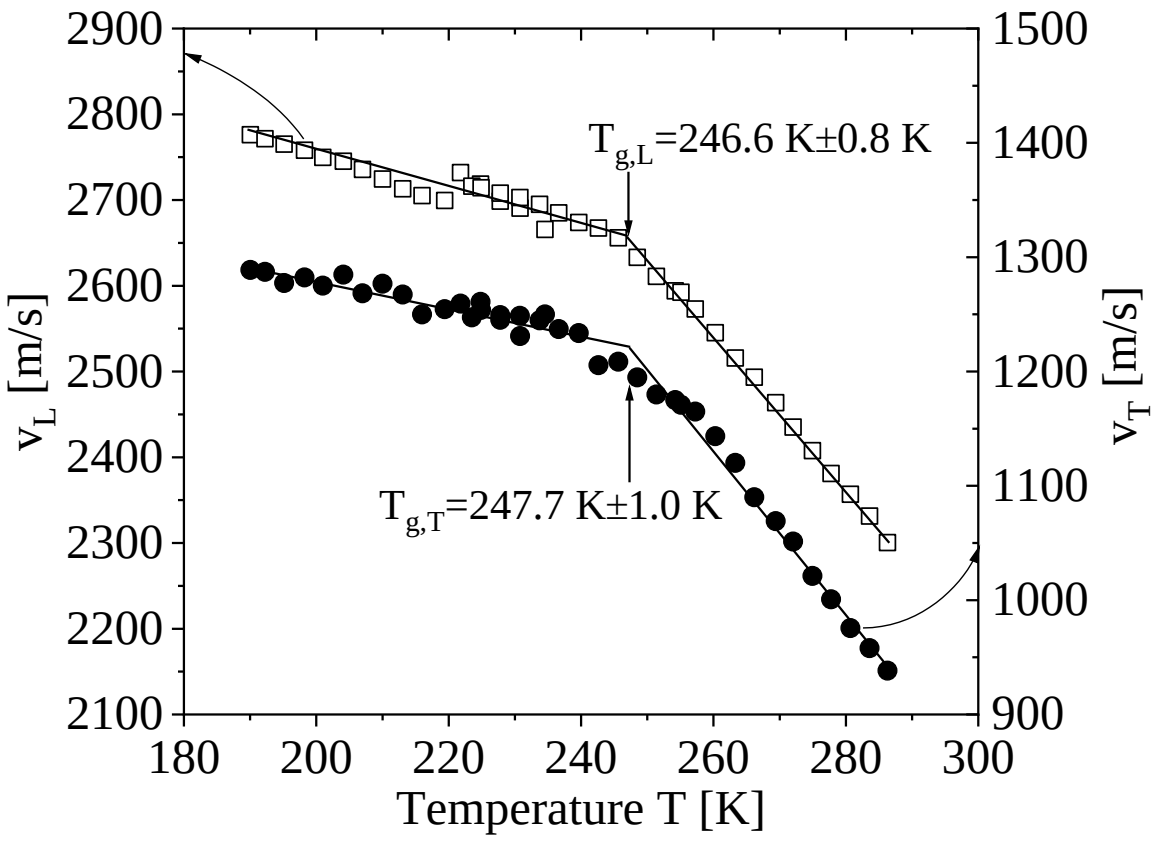

FIG 2. Clamped longitudinal and transversal sound velocities $v_L$ and $v_T$ as a function of temperature $T$ as measured via the temperature jump method. The kink indicates the quasi-static glass transition.

The fact that our model system DGEBA has a low quasi-static $T_g$ implies the advantage that the dynamic glass transition can be investigated at GHz frequencies without thermal degradation of the sample material.

Fig. 3 shows the behavior of the longitudinal modulus $c_L$ and the associated hypersonic losses $\Gamma^{90A}$ up to approx. 420 K. Due to the use of the temperature jump method, this measurement took more than 4 months. Since the mass density $\rho(T)$ changes little in the glass transition range compared to the elastic moduli, a constant value of 1000 kg/m$^3$ was used for the mass density $\rho$. At this point, it should be emphasized that the elastic modulus becomes a complex quantity as soon as the acoustic losses become significant $(c_L, \Gamma^{90A}) \Rightarrow (c'_{11}, c''_{22})$, (Voigt notation). Fig. 3 shows two distinct acoustic anomalies: i.) a pronounced maximum in the loss curve at 351.6 K and ii.) a relatively sharp kink in the modulus curve at $T_g = (246.9 \pm 1)\ K$. Within the scope of the statistical error, this value corresponds perfectly with those $T_g$'s obtained from the independent measurements shown in Fig. 2. The temperature $T_{g,dyn}$ specifies the dynamic glass transition of our model substance DGEBA. Within the Debye approximation (see for example, [24]), the dynamic $T_{g,dyn}$ is found at the temperature at which the condition $f/f_{rel} = 1$ is satisfied (maximum condition). According to the data given in Fig. 3 this condition is fulfilled for a relaxation frequency $f_{rel} = 4.5\ GHz$ at a temperature $T_{g,dyn} = 351.6\ K$. For the corresponding relaxation time at this very temperature one finds $\tau_{rel} = 35 \cdot 10^{-12} s$.

*Contact author: jan-kristian.krueger@ivw.uni-kl.de
+Contact author: andreas.klingler@ivw.uni-kl.de

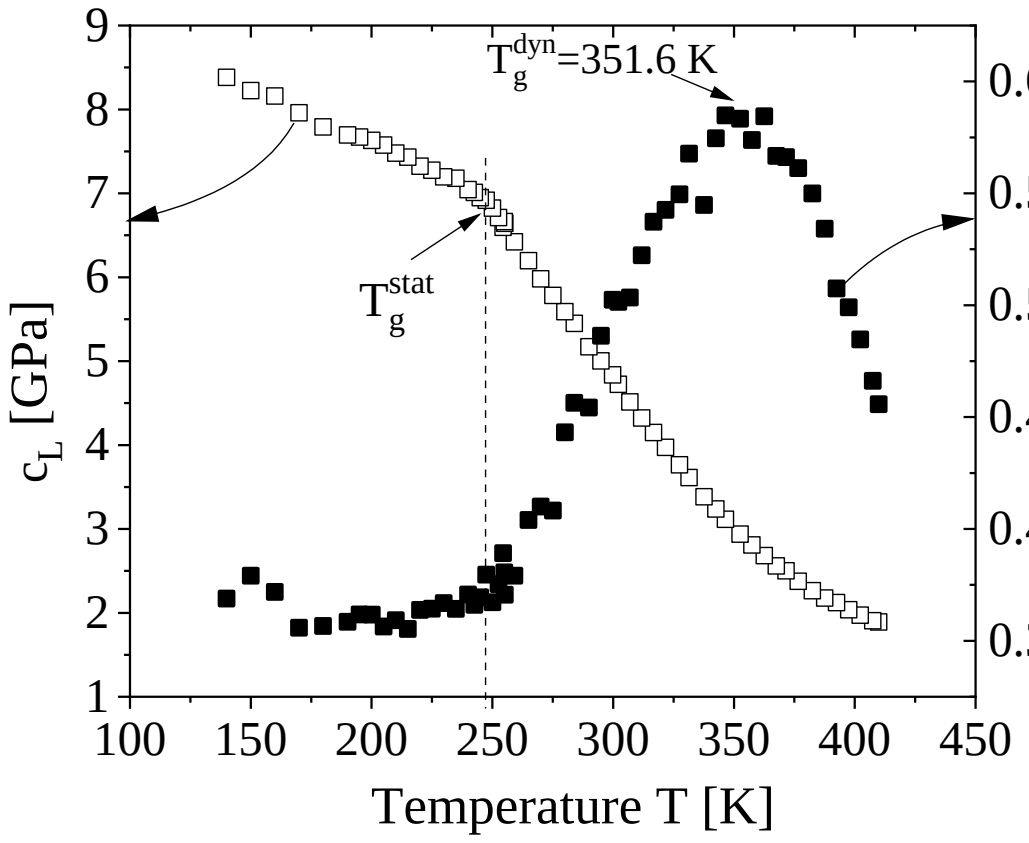

FIG 3. Longitudinal modulus $c_L$ and the associated hypersonic losses $\Gamma^{90A}$ of the model DGEBA upon cooling by temperature jumps. The data shows the dynamic $T_{g,dyn}$ and the quasi-static glass transition $T_{g,stat}$.

In one of our earlier publications [5], a Vogel-Fulcher-Tamann (VFT) behavior was predicted for the $\alpha$-relaxation behavior of DGEBA based on dielectric spectroscopy [25–27]. In Fig. 4, we have added this data to the relaxation frequency value that we measured in the high-temperature range using Brillouin spectroscopy.

If we tentatively connect the additional Brillouin data point with the high-frequency dielectric data points using a straight line, we find the typical VFT behavior: At high temperatures, the curved, low-temperature behavior of the VFT graph degenerates into an Arrhenius-like behavior. In this regard, the hypersonic curves shown in Fig. 2 behave as predicted in the literature.

However, there remain serious open questions. The most important question is probably: If the DGEBA probe freezes at approx. 351 K at Brillouin frequencies, what acoustic anomaly occurs at a temperature 100 K lower at $T_g \sim 247\ K$ where dynamically clamped modules $c_L^{\infty}$ are measured? In other words, what additional physical phenomenon produces a kink in the $c_L^{\infty}$-curve? Since we use Brillouin spectroscopy to measure the speed of sound rather than the modulus, Eq. 4b could provide a fairly simple answer: Since the mass density is in the denominator of Eq. 4b, it would appear that the mass density could cause the kink at $T_g$. However, since the kink in mass density and hypersonic velocity exhibit the same *convex* behavior, a contradiction arises. If the modulus $c_L$ shows no anomaly at $T_g$ but the mass density behaves convexly as a function of temperature, the temperature-dependent hypersonic velocity should behave *concavely* as a function of temperature. The opposite is the case. Therefore, an additional phenomenon must cause the kink in the hypersonic data.

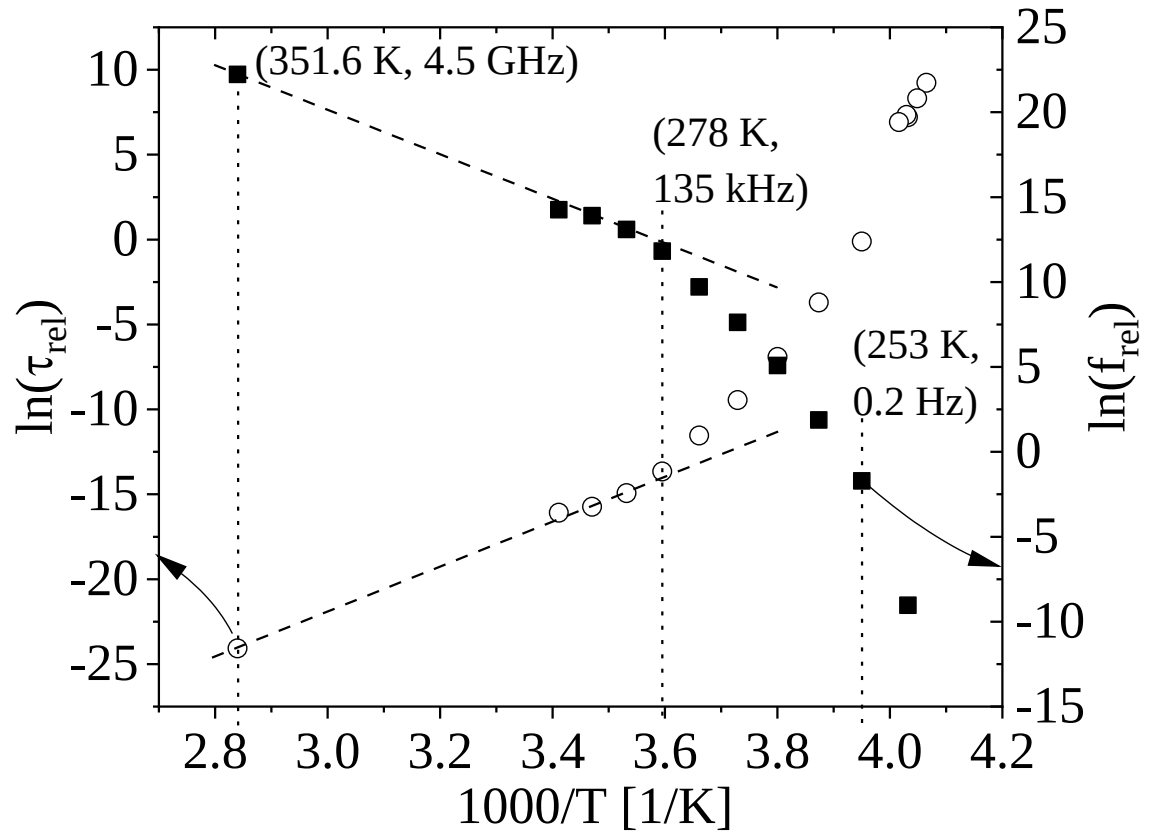

FIG 4. Vogel-Fulcher-Tammann diagram of the model DGEBA system based on combined data sets of dielectric measurements [5] and Brillouin spectroscopy (this work, 4.5 GHz).

There are further inconsistencies in connection with the hypersonic data presented above (Fig. 3). One of these inconsistencies concerns the extraordinary width of the acoustic damping maximum. It spans 100 K from the highest measured temperature to the thermal glass transition at $T_g$. If the $\alpha$-relaxation behavior presented in Fig. 4 is included in the assessment of the hypersonic attenuation behavior, it can be seen in Fig. 3 that significant hypersonic losses still exist at a temperature of 278 K, even though the $\alpha$-relaxation frequency has already dropped to 135 kHz. This raises the important question of the cause of hypersonic losses at temperatures at which the hypersonic modes should have been clamped long ago.

Furthermore, the nature of decay of the hypersonic loss curve at the quasi-static glass transition temperature is striking. It does indeed appear that the quasi-static (thermal) glass transition finally puts an end to the hypersonic losses at $T = T_{g,stat}$. Moreover, the sharp increase in the longitudinal modulus with decreasing temperature and approaching the quasi-static glass transition appears to contradict the Kramers-Kronig relation [28,29].

There is another extremely interesting observation that produces the hypersonic property at the thermal glass transition. As early as 2003, a generalized Cauchy relation (GCR) was observed in the thermal glass transition range in DGEBA and in some polymers [30]:

$$c_L - 3 \cdot c_T = A \qquad (5)$$

*Contact author: jan-kristian.krueger@ivw.uni-kl.de
+Contact author: andreas.klingler@ivw.uni-kl.de

Using temperature-jump measurements, i.e., by eliminating unwanted kinetic processes, the existence of the GCR for our model substance DGEBA was confirmed. Based on the data from Fig. 2, applying Eqs. 2, 3 and $\rho = 1000\ kg/m^3$, $c_L$ vs. $c_T$ is plotted in Fig. 5 and the GCR is calculated.

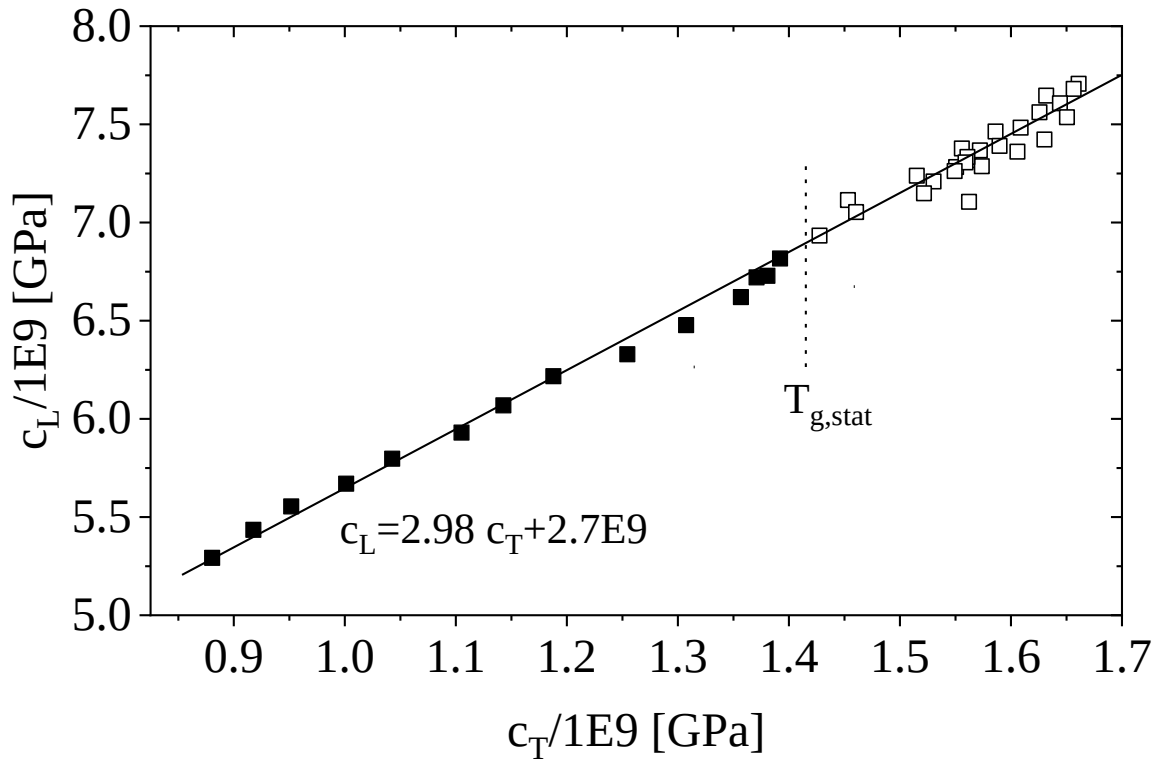

FIG 5. Generalized Cauchy relation across the quasi-static glass transition, whereas the sample temperature is a parameter, cf. Fig. 2. Bold squares: $T > T_g$, unfilled squares $T < T_g$

In Fig. 5, the linear $c_L(c_T)$ function crosses the quasi-static (thermal) glass transition at $T_g$ without any apparent anomaly, whereby the sample temperature $T$ is a parameter. This raises the question of why the kink-like thermal anomaly at $T_g$, which is clearly visible in the $c_L(T)$ and $c_T(T)$ curves in Fig. 2, remains hidden in the $c_L(c_T)$ representation and what consequences this has. To better understand the influence of the sample temperature on the GCR representation in Fig. 5, we fitted the underlying module curves from Fig. 2 with a generalized Taylor expansion:

$$c_L(T) = a_L - b_L \cdot T - c_L \cdot \left|T - T_{g,L}\right| \qquad (6a)$$

and

$$c_T(T) = a_T - b_T \cdot T - c_T \cdot \left|T - T_{g,T}\right| \qquad (6b)$$

Combining Eq. 5 with Eqs. (6a) and (6b) yields

$$A = a_L - b_L \cdot T - c_L \cdot \left|T - T_{g,L}\right| - 3a_T + 3b_T \cdot T + 3c_T \cdot \left|T - T_{g,T}\right| \qquad (7)$$

Taking into account that $T_{g,L} = T_{g,T}$ are identical for the longitudinal and the shear mode and that $A$ is a constant, Eq. 7 must apply to all temperatures, including $T = T_g$. It follows that the kink properties of the thermal glass transition given by the parameters $c_L$ and $c_T$ are lost, hence

$$A = (a_L - 3a_T) - (b_L - 3b_T) \cdot T_g \qquad (8)$$

and both acoustic modes are strongly coupled.

## IV. CONCLUSIONS

Combining all measured hypersonic properties for the model DGEBA, we here provide clear evidence for the existence of an independent static glass transition phenomenon that complements the dynamic glass transition properties. The generalized Cauchy relation supports this idea of an additional and independent glass transition phenomenon at $T_{g,stat}$.

## ACKNOWLEDGMENTS

The authors gratefully acknowledge the discussions with Martine Philipp to this work. The Brillouin data were created by Jan-Kristian Krüger at the Saarland University and later on at the University of Luxembourg.

*Contact author: jan-kristian.krueger@ivw.uni-kl.de
+Contact author: andreas.klingler@ivw.uni-kl.de

*Contact author: jan-kristian.krueger@ivw.uni-kl.de
+Contact author: andreas.klingler@ivw.uni-kl.de